\documentclass[lualatex]{pasj02}
\usepackage[switch,mathlines]{lineno}
\usepackage{url}
\usepackage{mathcomp}

\Received{$\langle$reception date$\rangle$}
\Accepted{$\langle$acception date$\rangle$}
\Published{$\langle$publication date$\rangle$}

\graphicspath{{./}{figure/}}

\begin{document}

\title{Classifying the derivatives of light curves for overcontact eclipsing binaries}

\author{Shinjirou Kouzuma}%

\altaffiltext{}{Faculty of Liberal Arts and Sciences, Chukyo University, 101-2 Yagoto-honmachi, Showa-ku, Nagoya, Aichi 466-8666, Japan}
\email{skouzuma@lets.chukyo-u.ac.jp}


\maketitle

\begin{abstract}
Recent studies indicate that the physical properties of eclipsing binaries can be extracted from the derivatives of their light curves. 
A classification scheme for the derivatives of light curves would be helpful for identifying key characteristics of eclipsing binaries. 
In this study, we propose a new classification method for the light curves of overcontact eclipsing binaries by using their derivatives. 
We synthesized 89,670 sample light curves of overcontact binaries and categorized them into five types on the basis of their first to fourth derivatives. 
For each type, we examined the statistical distributions of four parameters: the mass ratio, orbital inclination, fill-out factor, and eclipse obscuration. 
Their distributions demonstrated that parameter values exhibit certain trends depending on the classified types. 
With the proposed classification method, general properties of overcontact binaries can be understood, providing a foundation for further detailed analysis. 
\end{abstract}

\begin{figure*}[]
 \begin{center}
  \includegraphics[width=0.44\textwidth]{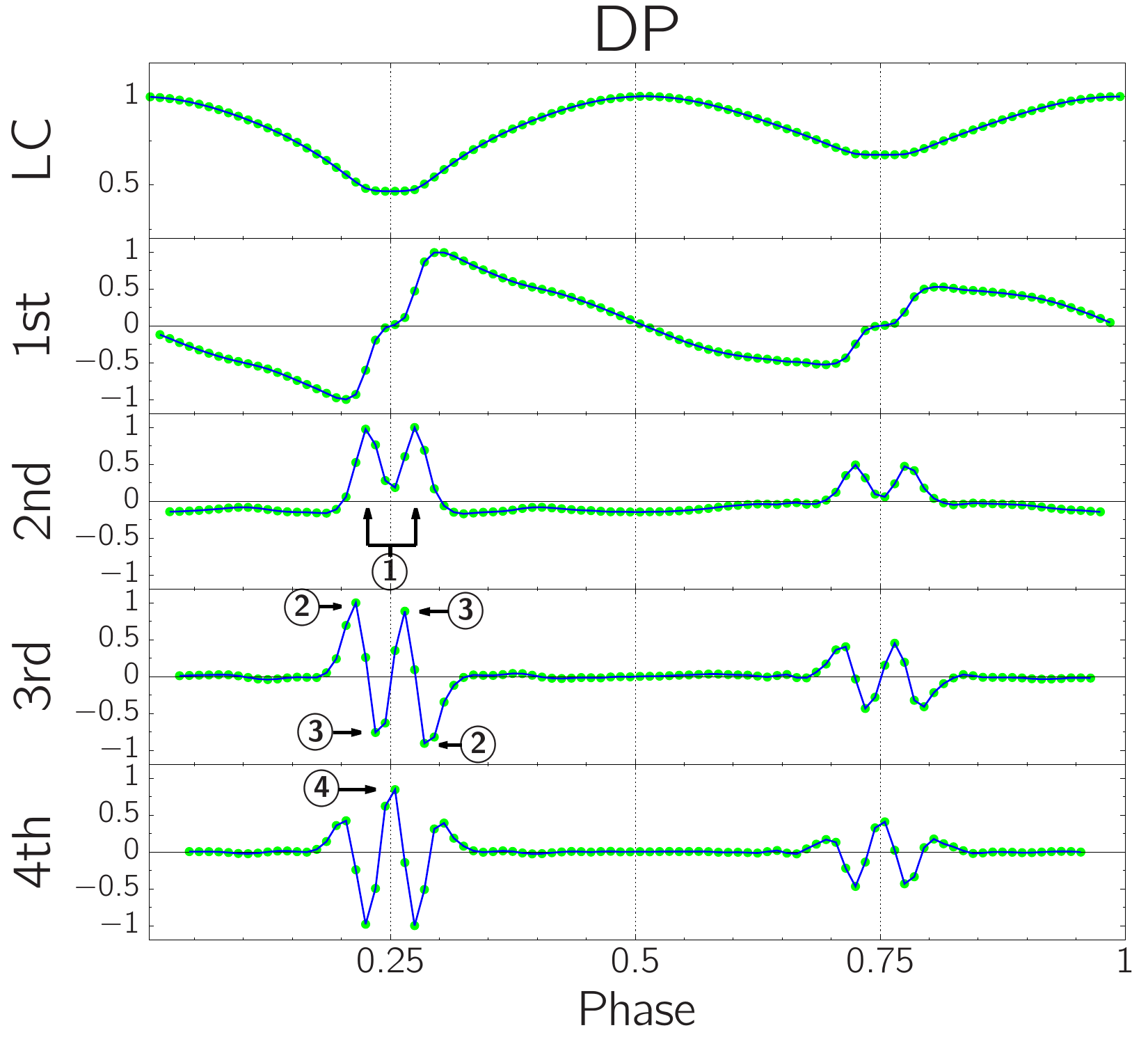}
  \includegraphics[width=0.44\textwidth]{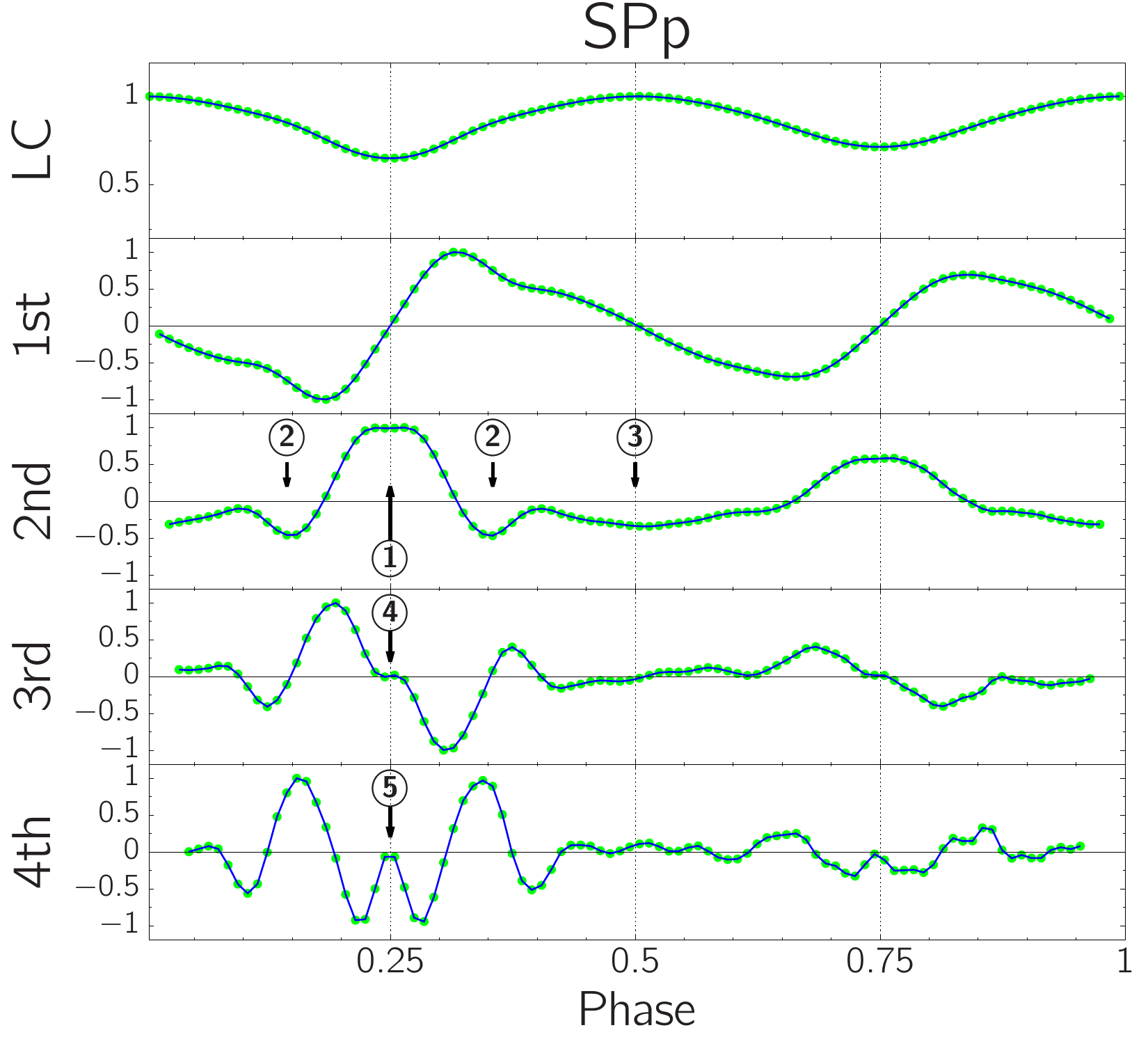}
  \includegraphics[width=0.44\textwidth]{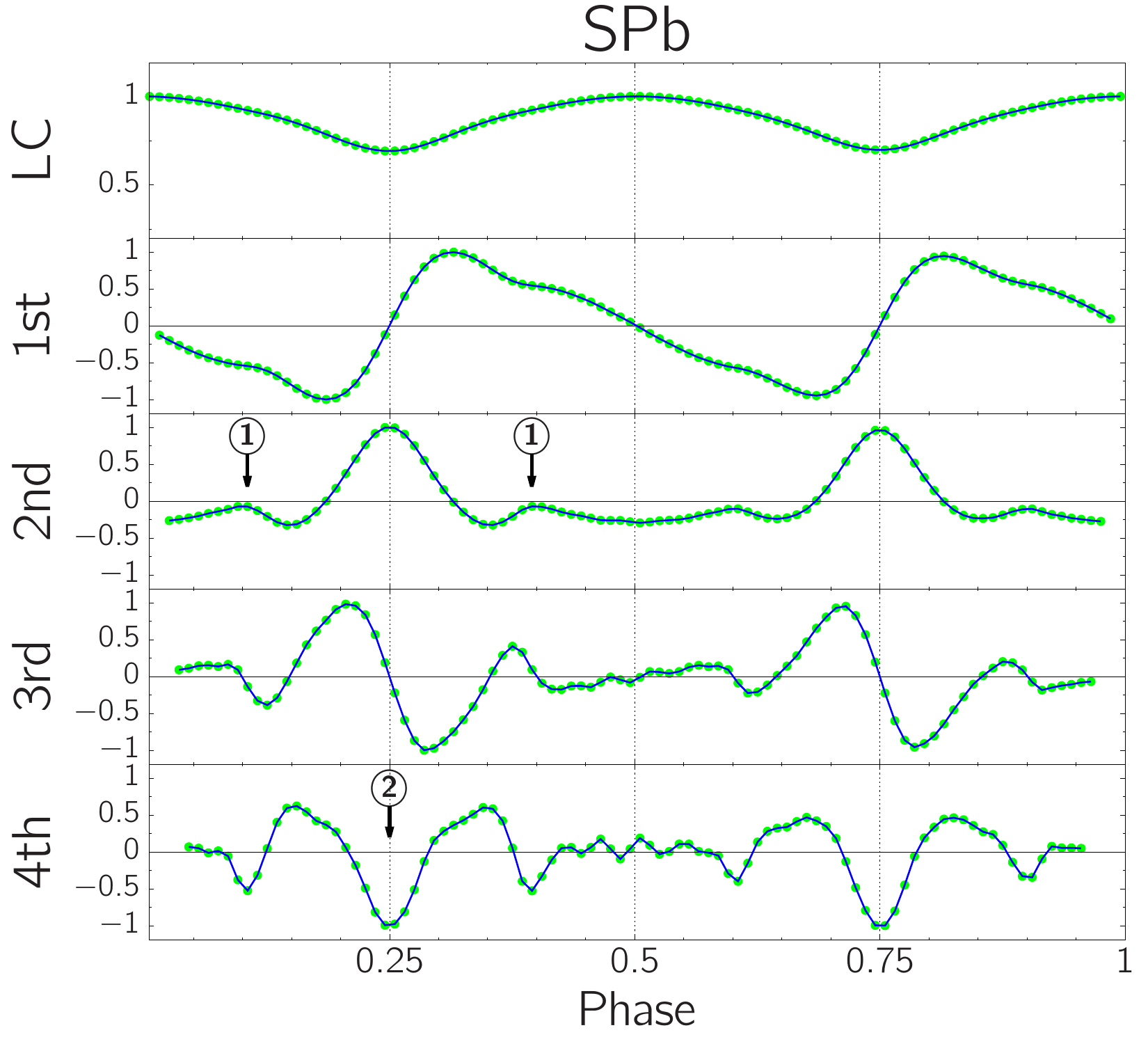}
  \includegraphics[width=0.44\textwidth]{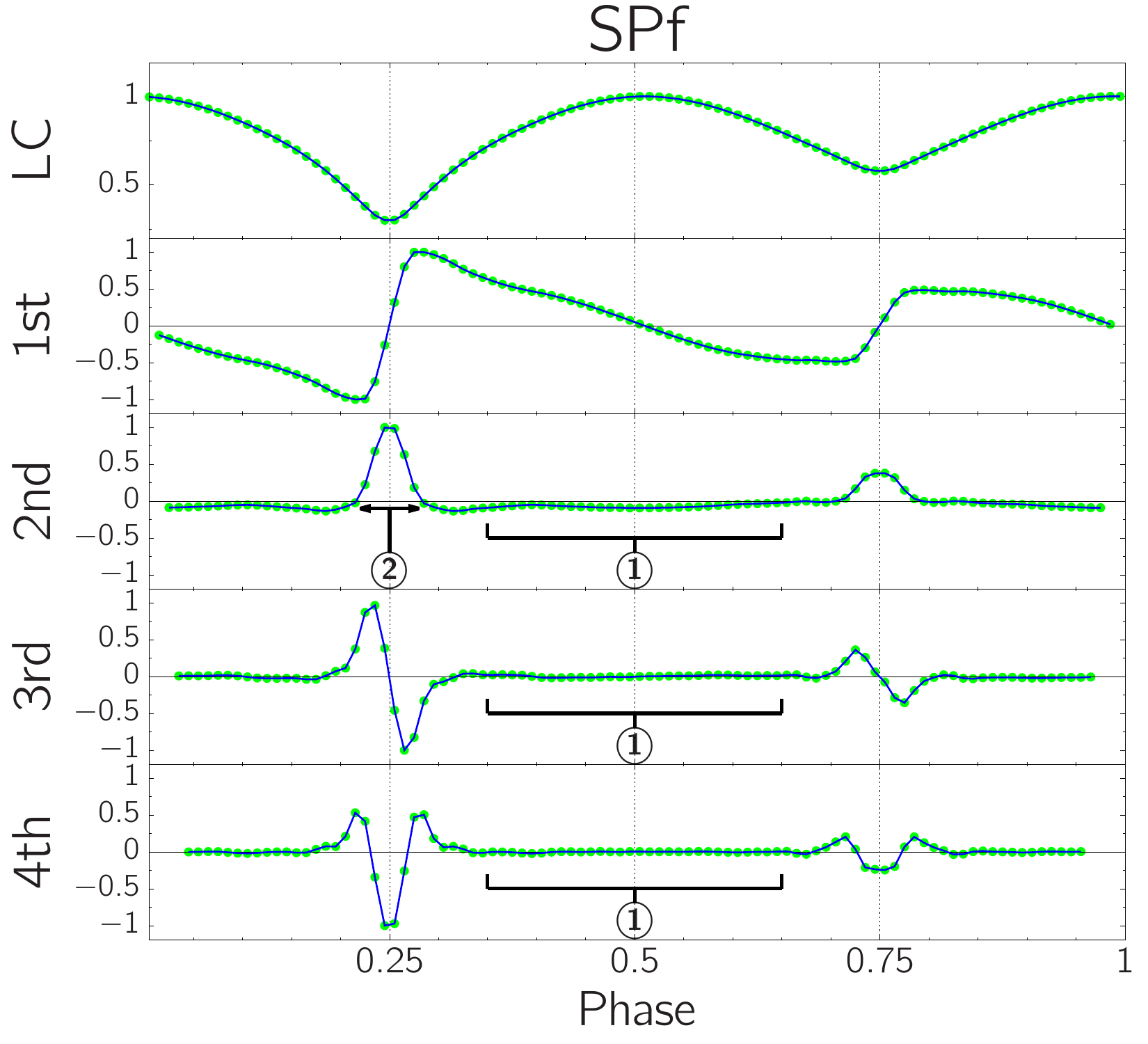}
  \includegraphics[width=0.44\textwidth]{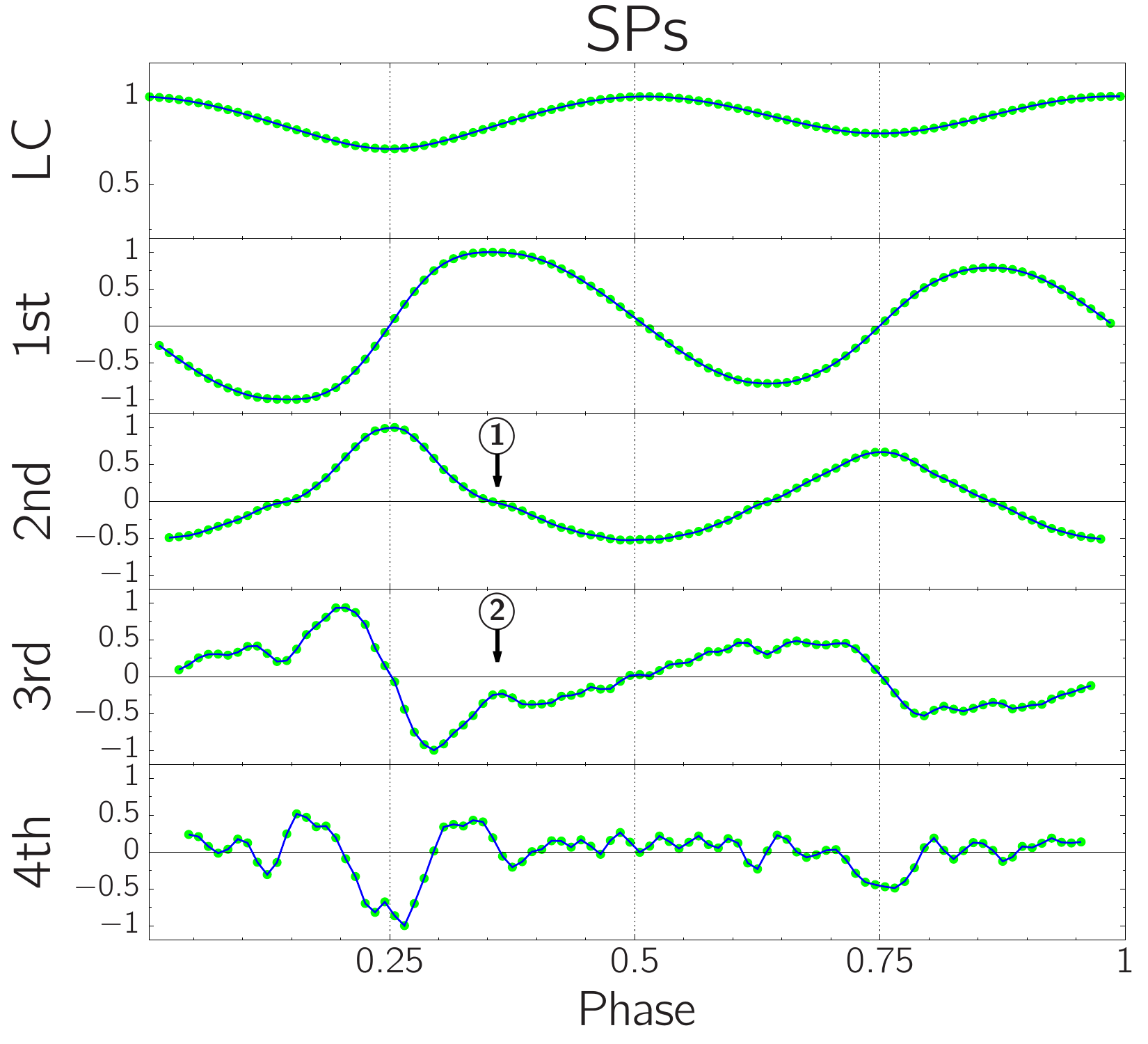}
 \end{center}
 \caption{Light curves and their first to fourth derivaives (from top to bottom) of the representative sample binaries for the proposed types. 
			Each derivative is normalized to unity. 
			Features described in the text are marked with numbers. 
		 {Alt text: Five panels illustrate example light curves and their derivatives for each of the proposed types. X axes show the phase from 0 to 1. }
 \label{fig_types}}
\end{figure*}
\section{Introduction}
The light curves (LCs) of eclipsing binary stars embody essential information about the nature of these systems. 
Classifying LCs is a fundamental and crucial work, providing a basic understanding that guides further detailed analysis. 
This initial categorization allows us to identify key characteristics of the binary systems quickly. 

The EW, EB ($\beta$-Lyrae), and EA (Algol) types are widely used classifications for the LCs of eclipsing binaries (e.g., \cite{Samus2004-catalogue}). 
This classification is based on the shapes of the LCs, and the differences in these shapes are partly related to the physical shapes of the component stars. 
Furthermore, these three types are also associated with morphological types proposed by \citet{Kopal1955-AnAp} \citep{Kallrath2009-book}, in which close binaries are classified on the basis of Roche geometry: contact, semi-detached, and detached systems. 
The LCs of EW type are often indicative of overcontact binary systems. 
An overcontact binary has two component stars with overfilled Roche lobes and their surfaces are combined. 
Such a shape produces ellipsoidal variation and is responsible for continuous light variation observed in the LCs of EW-type systems. 
Additionally, EW systems are further divided into A- and W-subtypes \citep{Binnendijk1970-VA}. 
These types are distinguished by whether the primary (deeper) minimum in a LC is caused by a transit (A-subtype) or by an occultation (W-subtype). 
Observational studies have found differences in physical properties between the two subtypes: 
e.g., A-subtype systems, compared to W-subtype systems, have longer orbital periods \citep{Yamasaki1975-ApSS, Smith1984-QJRAS, Gazeas2006-MNRAS}, smaller mass ratios \citep{Smith1984-QJRAS}, and earlier spectral types \citep{Rucinski1974-AcA, Webbink2003-ASPC}. 
Therefore, an appropriate classification for the LCs of eclipsing binaries allows us to obtain essential and general information about the binary systems. 

Recent studies have explored the physical properties of eclipsing binaries by using the derivatives of LCs. 
Eclipses have been demonstrated to have typical signatures in the derivatives of LCs \citep{IJspeert2021-AA, IJspeert2024-AA}. 
\citet{Kouzuma2023-ApJ} proposed a simple method for estimating the mass ratios of overcontact binaries using the first to third derivatives of LCs. 
These studies indicate that the derivatives of LCs have the potential for obtaining insights into the nature of eclipsing binaries. 
Therefore, a classification scheme based on the derivatives of LCs would be helpful for identifying key characteristics of binaries quickly. 
However, no such classification scheme has been reported. 

In this paper, we propose a new classification scheme for the LCs of overcontact binaries, which is based on the derivatives of the LCs. 
We also investigate the associations of the proposed types with four fundamental parameters. 
Section \ref{Sec_LC} introduces sample LCs for overcontact systems, and section \ref{Sec_Classification} describes a method for classifying them. 
In section \ref{Sec_Result}, we present five types of LCs classified based on the shapes of the derivatives and describe the characteristics of the derivatives for each type. 
Additionally, we discuss the associations of the proposed types with binary parameters. 
The summary of this study is in section \ref{Sec_Summary}.

\section{Sample LCs}\label{Sec_LC}
We synthesized LCs for overcontact binaries using the PHOEBE 2.4 code \citep{Prsa2016-ApJS, Conroy2020-ApJS}. 
In this study, we adopted the following parameter ranges and steps: the mass ratio $q=M_\mathrm{s}/M_\mathrm{p}=0.05$--$0.95$ (in steps of 0.1), orbital inclination $i=30\tcdegree$--$90\tcdegree$ ($1\tcdegree$), fill-out factor $f=0.2$--$0.8$ (0.3), primary- and secondary-star temperatures $T_\mathrm{p} (T_\mathrm{s})=$ 4000--10000 (1000) K. 
Here, indices `p' and `s' refer to primary and secondary, respectively. 
Throughout this paper, the primary star is defined as the more massive component of the binary; the mass ratio does not exceed 1. 
The gravity-darkening coefficient was set either to 0.32 or 1, depending on whether the star's temperature was lower or higher than 6600 K, respectively. 
The fluxes were computed at every 0.001 step in phase. 
Finally, a total of 89,670 LCs were generated.

\section{Classification}\label{Sec_Classification}
The goal of this paper is to propose a new classification scheme based on the derivatives of LCs, which enables the understanding of the general properties of overcontact binaries from the classified LC types. 
We classify the LCs of overcontact binaries on the basis of the derivatives of each LC with respect to time. 
Using the synthesized LCs introduced in section \ref{Sec_LC}, we computed the numerical derivatives of each LC up to the fourth order. 
This work utilizes the second-order central difference for the mean fluxes, each data point of which is 100 phase bins combined.
We find a higher-order derivative for each by repeating numerical differentiation. 

For this classification, we analyzed the shapes of their derivative curves, which were visually inspected and compared to identify common features. 
Our classification was performed by grouping the LCs according to the similarities observed in their derivative behaviors, particularly focusing on the local maximum and minimum points, slopes, and inflection points in the curves. 
On the basis of these observed similarities, we categorized the LCs into distinct groups. 
Through this process, we minimized the number of categories to ensure the classification was as simple and efficient as possible.

\begin{figure}
\begin{center}
  \includegraphics[width=0.45\textwidth]{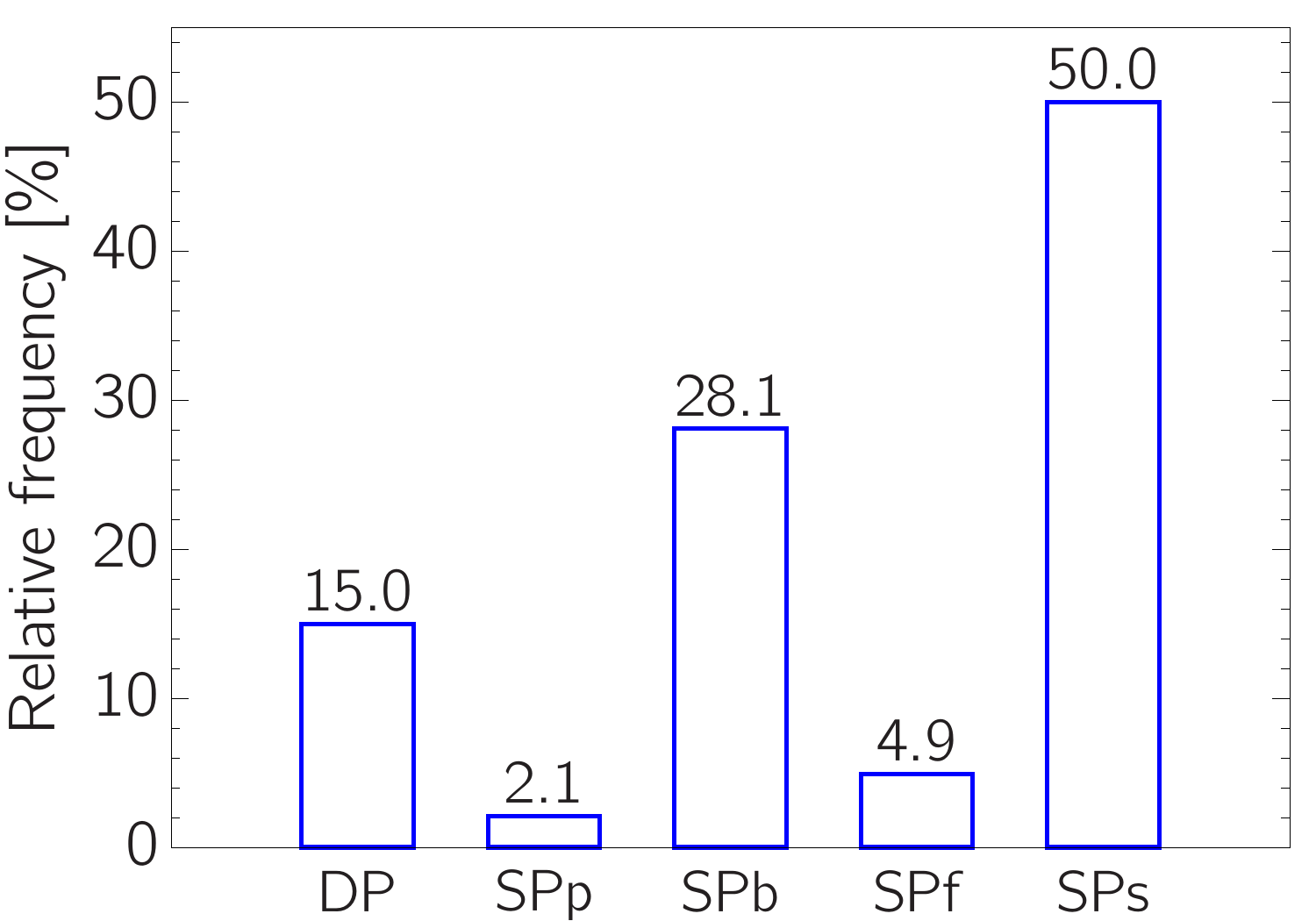}
 \end{center}
 \caption{Fractions of classified binaries for each type relative to the total. 
		Each value is shown above the corresponding bar. 
		{Alt text: Relative frequency bar chart. }
		\label{fig_types_hist}}
\end{figure}

\begin{figure*}
\begin{center}
  \includegraphics[width=0.98\textwidth]{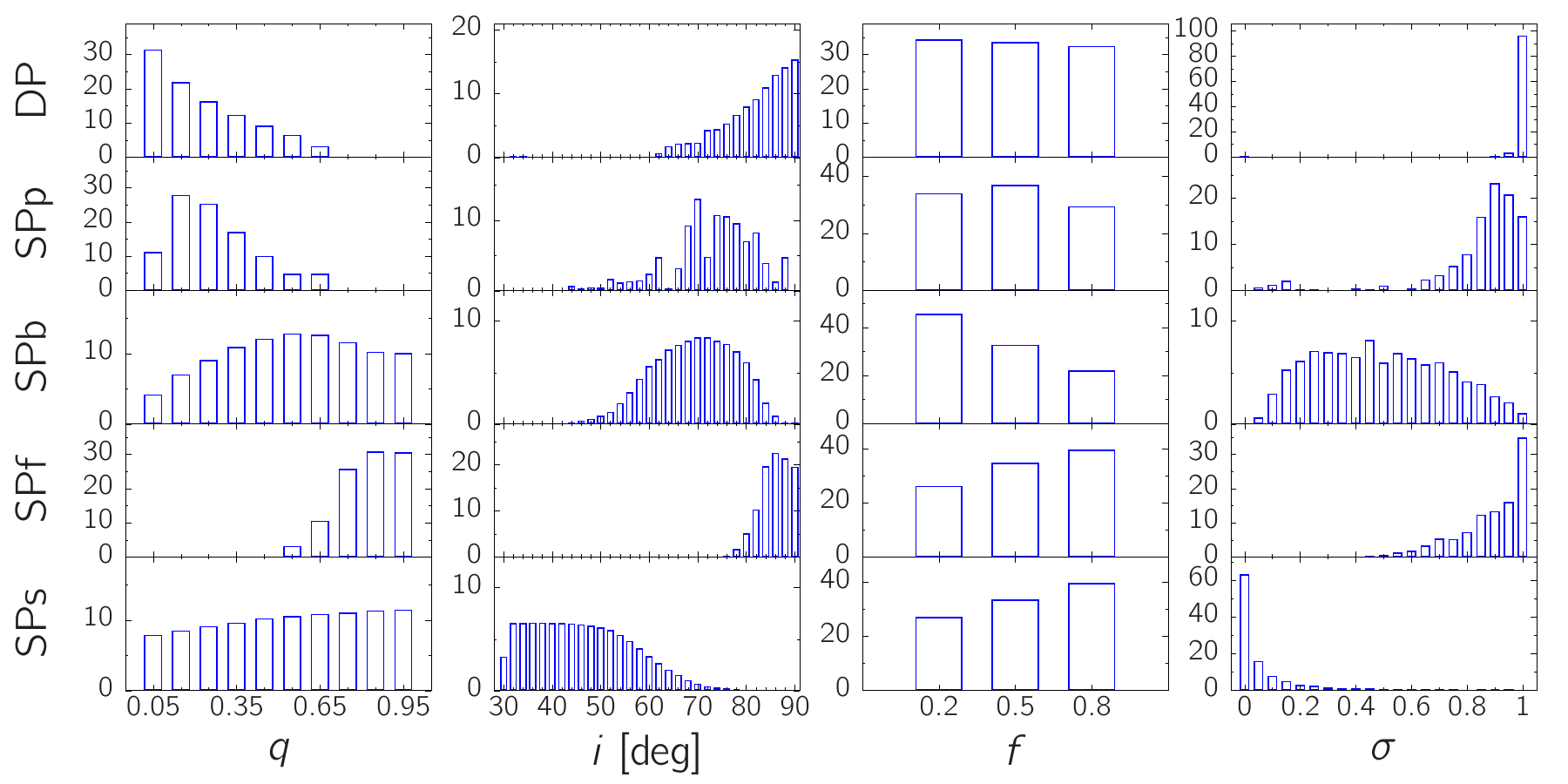}
 \end{center}
 \caption{Histograms of four binary parameters (the mass ratio, orbital inclination, fill-out factor, and eclipse obscuration) for each classified binary. 
			The vertical axes refer to the relative frequency presented in percentages. 
			{Alt text: Relative frequency histograms. x axes show the mass ratio from 0.05 to 0.95, orbital inclination from 30 to 90 degrees, fill-out factor from 0.2 to 0.8, and eclipse obscuration from 0 to 1. }
			\label{fig_types_parms_hist}}
\end{figure*}
\section{Results and Discussion}\label{Sec_Result}
We determined that it is appropriate to classify the synthesized LCs into five groups on the basis of the geometry of their derivatives: DP, SPp, SPb, SPf, and SPs. 
Figure \ref{fig_types} shows examples of LCs and their derivatives for representative sample binaries corresponding to these five types. 
Markers with numbers in the figure indicate the features described in the text; they are labeled using the names of the proposed five types, followed by a hyphen and number (e.g., DP-1). 
In figure \ref{fig_types_hist}, we show the fractions of our proposed types for the synthesized LCs. 
Additionally, figure \ref{fig_types_parms_hist} summarizes the statistical distributions of four parameter values for each type: the mass ratio ($q$), orbital inclination ($i$), fill-out factor ($f$), and eclipse obscuration ($\sigma$). 
Throughout this paper, the eclipse obscuration is defined as the maximum value of the fraction of the smaller star's surface area covered by the larger star during the eclipse. 
For instance, $\sigma$ reaches a value of 1 when the smaller star is fully occulted by the larger star. 

In the following subsections, we first describe the key features of the LC derivatives for each of the five types. 
We then discuss the physical properties of each type on the basis of the statistical distributions of the four parameters. 

\subsection{DP: double peak in the second derivative}\label{Sec_DP}
Two local maxima (hereafter referred to as "double peak") appear symmetrically around the eclipse time in the second derivative of a LC (DP-1), as seen in figure \ref{fig_types}. 
In the third derivative, a pair of features is present: a local maximum just before the eclipse time and a local minimum just after it (DP-2). 
If the double peak is clearly distinguishable, another local minimum-maximum pair may appear in the third derivative (DP-3). 
In this situation, a local maximum will be present at the eclipse time in the fourth derivative (DP-4). 
Note that the peak of DP-4 may sometimes split into two. 
Among our synthesized LCs, nearly 15\% exhibited double peaks (see figure \ref{fig_types_hist}). 
Almost all of the synthesized LCs with such double peaks were those of binaries exhibiting total-annular eclipses (figure \ref{fig_types_parms_hist}); 
the remaining few percent exhibited double peaks that were difficult to distinguish from single peaks. 
This indicates that the presence of a double peak can be a criterion for identifying total-annular eclipses. 

In figure \ref{fig_types_parms_hist}, over 90\% of the DP systems have $q<0.6$ and $i>70^\circ$. 
The histograms for $q$ and $i$ display that their frequencies gradually increase as the mass ratio decreases and the orbital inclination increases. 
These trends indicate that overcontact binaries with lower mass ratios and higher inclinations tend to show DP-type LCs, which result in almost all DP systems exhibiting total-annular eclipses. 
Note that \citet{Kouzuma2023-ApJ} proposed a method for estimating the mass ratios of overcontact binaries with double peaks appearing in the second derivatives of their LCs. 

\subsection{SP: single peak in the second derivative}\label{Sec_SP}
In the second derivative of a LC, a local maximum (single peak) commonly appears at the time of an eclipse instead of a double peak (DP-1). 
Below, we further subdivide this type into four categories. 
\subsubsection{SPp: SP with a peak in the fourth derivative}\label{Sec_SPp}
In the fourth derivative of a LC, a local maximum (SPp-5) is observed at an eclipse time, which arises due to a shallow slope in its third derivative (SPp-4). 
In the second derivative, the shape of a peak around the eclipse time (SPp-1) tends to be flatter or, in some cases, has an extremely shallow depression. 
This flattening becomes more pronounced as the peak in the fourth derivative (SPp-5) increases in height. 
Additionally, as the SPp-5 peak diminishes, the eclipse obscuration tends to be relatively small. 
In our synthesized LC samples, this type is the least common, comprising approximately 2\%, as shown in figure \ref{fig_types_hist}. 

If a LC has an inadequately small number of data points or insufficient photometric precision, a peak like SPp-5 may appear in SPb or SPs-type LCs even when it does not actually exist, making them easily mistaken for SPp. 
In this situation, such a spurious peak is fairly small in most cases. 
In addition, if the local minimum around phase 0.5 in the second derivative (SPp-3) is distinctly deeper than those before and after the eclipse (SPp-2), its LC is more likely not to belong to the SPp type. 
Almost all of the LCs with $\sigma \sim 0.1$--$0.2$ (see figure \ref{fig_types_parms_hist}) are misclassified as SPp due to the detection of spurious SPp-5 peaks. 
However, a visual inspection reveals that they actually belong to the SPb or SPs categories. 

SPp systems tend to have low mass ratios (see figure \ref{fig_types_parms_hist}); within the range $q<0.6$, their frequency increases as the mass ratio decreases, peaking at a mass ratio of 0.15. 
Only about 10\% of the SPp systems have mass ratios below 0.1. 
Although the SPp systems have various inclinations ranging between $60^\circ$ and $90^\circ$, their eclipse obscurations remain notably high, with nearly 95\% exceeding 0.7. 
The SPp-type LCs with $0.6<\sigma<0.8$ are systems having $q<0.3$. 
SPp systems are likely to have low mass ratios, as DP systems have, but most of their orbital inclinations are insufficiently high to result in total-annular eclipses. 

\subsubsection{SPb: SP with small bumps in the second derivative}
The LCs and their derivatives of SPb systems exhibit shapes that are very similar to those of SPp systems. 
However, the SPb systems display no peaks in the fourth derivative at the eclipse time (SPb-2), which are observed in the SPp-type LCs. 
Two small bumps (local maxima) remain appearing symmetrically before and after the eclipse (SPb-1). 
These bumps can be detected by finding zero values in the third derivative. 
Approximately 28\% of the synthesized LCs were classified as SPb type (figure \ref{fig_types_hist}). 

SPb systems have wide ranges of mass ratios, orbital inclinations, and eclipse obscurations (figure \ref{fig_types_parms_hist}). 
Whereas the frequency decreases with decreasing mass ratio below $q \sim 0.6$, it is relatively uniform above $q \sim 0.6$. 
The orbital inclinations of the SPb systems is greater than $50^\circ$ but hardly exceed $85^\circ$, 
resulting in a low proportion of systems exhibiting high eclipse obscurations. 
Only 1\% of the SPb systems exhibit total-annular eclipse, and nearly 95\% have eclipse obscurations within $0.1<\sigma<0.9$. 
The frequency decreases with increasing fill-out factor. 
The SPb-type LCs may be further subdivided. 

\subsubsection{SPf: SP with a flat shape outside of eclipse}
In the second to fourth derivatives, their shapes remain mostly flat with little variation outside the vicinity of the eclipse time (SPf-1). 
This flatness is caused by the rapid changes in luminosity around the eclipse time, where the rate of change is several times higher than that observed in systems of other types. 
In the flat regions, the amount of change is generally less than 10\% of the maximum variation in the corresponding derivative. 
SPf-type LCs typically exhibit narrower peaks in the second derivative than those of other types (SPf-2). 
Only approximately 5\% of the synthesized LCs were classified as this type. 

The SPf systems tend to have high mass ratios ($q>0.6$) and high inclinations ($i>80^\circ$). 
Approximately 35\% of the SPf systems exhibited total-annular eclipses ($\sigma \sim 1$), and almost all (nearly 95\%) had eclipse obscurations above 0.7. 
The frequency increases with increasing fill-out factor. 
Overcontact systems with high mass ratios, indicating that the sizes of the two component stars are similar, and eclipses with high obscurations including total-annular ones should be classified as SPf type. 

\subsubsection{SPs: SP with a smooth curve in the second derivative}
The two bumps in the second derivative of the SPb-type LC (SPb-1) are absent in SPs-type LCs (see SPs-1). 
This absence results in the local maximum in the third derivative being negative (SPs-2). 
The second derivative exhibits smooth variation throughout the phase. 
Nearly 50\% of the synthesized LCs were classified as this type. 

A distinguishing feature of the SPs systems is their extremely low eclipse obscurations. 
Unlike other types, their eclipse obscurations hardly exceed 0.4, and approximately 95\% have $\sigma<$ 0.2. 
In particular, 12,544 (28\%) of the SPs systems exhibited extremely low eclipse obscurations of less than 0.1\% (i.e., $\sigma<10^{-3}$). 
This characteristic is likely attributed to the tendency of SPs systems to have low orbital inclinations (see figure \ref{fig_types_parms_hist}): 
almost all of the SPs systems have $i<70^\circ$. 
The frequency increases with increasing fill-out factor. 

As mentioned in section \ref{Sec_SPp}, SPs-type LCs display peaks in the fourth derivative similar to SPp-5 in some cases. 
SPs systems tend to have low orbital inclinations and small eclipse obscurations, which lead to small variability amplitudes in LCs and, consequently, noisy derivative curves. 
This can result in the appearance of such spurious peaks.

\begin{figure}
\begin{center}
  \includegraphics[width=0.45\textwidth]{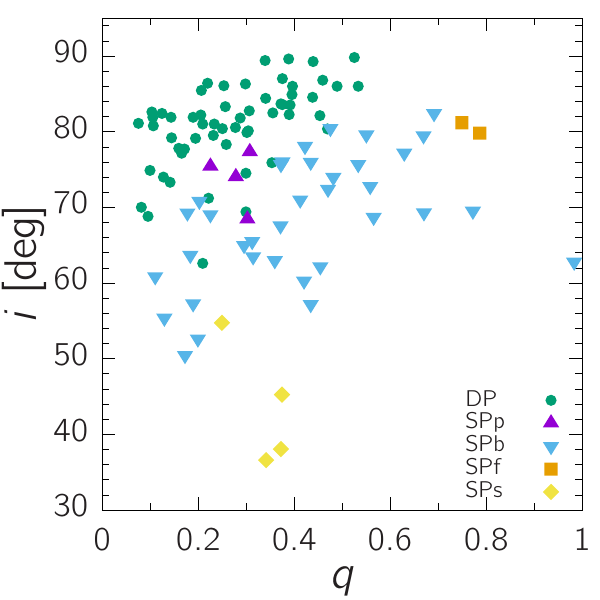}
 \end{center}
 \caption{catter plot of the orbital inclination angle against the mass ratio.  
		{Alt text: The x-axis shows the mass ratio ranging from 0 to 1, and the y-axis shows the orbital inclination angle ranging from $30^\circ$ to $90^\circ$. }
		\label{fig_q-i}}
\end{figure}
\subsection{Application to real binary data}
We applied our proposed classification method to real binary data with spectroscopic mass ratios. 
In the catalog by \citet{Latkovic2021-ApJS}, 159 W UMa systems have entries of spectroscopic mass ratios. 
We searched the TESS and Kepler archival data for their LCs using the Python package Lightkurve \citep{Lightkurve2018-code}; LCs for 127 binaries were found. 
When two or more LCs were found for a binary, the better-quality one was selected in such a manner that its derivatives were less noisy and smoother than the other one. Apparent outliers in the LCs were removed in advance. 
We then derived the derivatives of the selected light curves. 
Applying the proposed method to the LCs, we classified them as follows: 57 DP, 4 SPp, 35 SPb, 2 SPf, and 4 SPs. 
The remaining 25 LCs were ruled out because they display noisy curves in their derivatives. 
Such noisiness is likely due to insufficient time resolution or variability of the component stars, such as pulsations.

Table \ref{tab_Classify} shows the classified types of 102 binaries along with their mass ratio, orbital inclination, and fill-out factor. 
Figure \ref{fig_q-i} presents a scatter plot of the orbital inclination against the mass ratio for the 102 binaries. 
For each type, the distribution on the plot is consistent with the physical properties described in sections \ref{Sec_DP} and \ref{Sec_SP}.

\begin{longtable}{lllll}
  \caption{Classified types and three fundamental parameters of binaries with entries of spectroscopic mass ratios in \citet{Latkovic2021-ApJS}. 
		   The parameter values were extracted from Latkovi{\'c}'s catalog. 
	}\label{tab_Classify}  
\hline\noalign{\vskip3pt} 
  Name & Type & Mass ratio & Inclination & Fill-out factor \\   [2pt] 
\hline\noalign{\vskip3pt} 
\endfirsthead      
\hline\noalign{\vskip3pt} 
  Name & Type & Mass ratio & Inclination & Fill-out factor \\  [2pt] 
\hline\noalign{\vskip3pt} 
\endhead
\hline\noalign{\vskip3pt} 
\endfoot
\hline\noalign{\vskip3pt} 
\multicolumn{2}{@{}l@{}}{\hbox to0pt{\parbox{160mm}{\footnotesize
\hangindent6pt\noindent
}\hss}} 
\endlastfoot 

1SWASP J093010.78+533859.5  &  DP  &  0.397  &  86  &  0.173211        \\
1SWASP J034501.24+493659.9  &  SPb  &  0.421  &  60.3  &  0.249087     \\
1SWASP J160156.04+202821.6  &  SPb  &  0.67  &  79.5  &  0.129425      \\
44 Boo  &  SPb  &  0.559  &  72.8  &  0                                \\
AA UMa  &  SPb  &  0.551  &  79.61  &  0.143285                        \\
AC Boo  &  DP  &  0.299  &  86.3  &  0.046157                          \\
AD Phe  &  SPb  &  0.376  &  76.05  &  0.163683                        \\
AG Vir  &  DP  &  0.341  &  84.4  &  0.167                             \\
AO Cam  &  SPb  &  0.435  &  76  &  0.17497                            \\
AQ Psc  &  SPb  &  0.226  &  69.06  &  0.301475                        \\
AQ Tuc  &  DP  &  0.354  &  75.89  &  0.36645                          \\
AU Ser  &  SPb  &  0.692  &  82.43  &  0.0406                          \\
AW UMa  &  DP  &  0.076  &  81.1  &  0.487                             \\
BD+422782  &  SPb  &  0.482  &  74  &  0.374802                       \\
BI CVn  &  SPb  &  0.413  &  71.01  &  0.309188                        \\
BN Ari  &  DP  &  0.392  &  83.52  &  0.125475                         \\
BO Ari  &  DP  &  0.19  &  81.9  &  0.493753                           \\
BO CVn  &  DP  &  0.34  &  89.4  &  0.82537                            \\
BV Eri  &  DP  &  0.3  &  74.5  &  0.321969                            \\
BX Dra  &  DP  &  0.288  &  81.8  &  0.604006                          \\
BX Peg  &  DP  &  0.376  &  87  &  0.175006                            \\
CC Com  &  DP  &  0.526  &  89.8  &  0.155827                          \\
CK Boo  &  SPb  &  0.111  &  60.87  &  0.946331                        \\
CN Hyi  &  SPb  &  0.184  &  63.7  &  0.428118                         \\
CT Eri  &  DP  &  0.3  &  69.4  &  0.690826                            \\
DK Cyg  &  DP  &  0.307  &  82.75  &  0.448098                         \\
DU Boo  &  DP  &  0.234  &  81  &  0.508353                            \\
DY Cet  &  DP  &  0.356  &  82.48  &  0.264273                         \\
EE Cet  &  SPb  &  0.315  &  63.5  &  0.21834                          \\
EF Boo  &  SPb  &  0.534  &  75.7  &  0.248056                         \\
EF Dra  &  DP  &  0.16  &  77.8  &  0.467413                           \\
EL Aqr  &  SPb  &  0.203  &  70.84  &  0.441778                        \\
$\epsilon$ CrA  &  DP  &  0.128  &  74  &  0.267834                           \\
EQ Tau  &  DP  &  0.439  &  84.54  &  0.092754                         \\
ET Leo  &  SPs  &  0.342  &  36.6  &  0.546359                         \\
EZ Hya  &  DP  &  0.257  &  83.3  &  0.34262                           \\
FG Hya  &  DP  &  0.104  &  82.6  &  0.697789                          \\
FI Boo  &  SPs  &  0.373  &  38.05  &  0.487448                        \\
FN Cam  &  DP  &  0.222  &  71.2  &  0.88                              \\
FP Boo  &  DP  &  0.096  &  68.8  &  0.393701                          \\
FT Lup  &  DP  &  0.44  &  89.26  &  0.133483                          \\
FT UMa  &  SPb  &  0.984  &  62.8  &  0.153086                         \\
FU Dra  &  DP  &  0.251  &  80.4  &  0.267468                          \\
GM Dra  &  DP  &  0.21  &  62.6  &  0.400341                           \\
GR Vir  &  DP  &  0.106  &  81.9  &  0.931822                          \\
HI Dra  &  SPs  &  0.25  &  54.74  &  0.229657                         \\
HI Pup  &  DP  &  0.206  &  82.2  &  0.202567                          \\
HV Aqr  &  DP  &  0.145  &  79.19  &  0.568318                         \\
HV UMa  &  SPb  &  0.19  &  57.3  &  0.77                              \\
II UMa  &  DP  &  0.172  &  77.7  &  0.868578                          \\
KIC 10618253  &  DP  &  0.125  &  82.4  &  0.93                        \\
LO And  &  DP  &  0.305  &  80.1  &  0.401383                          \\
LS Del  &  SPs  &  0.375  &  45.25  &  0.196555                        \\
MW Pav  &  DP  &  0.22  &  86.39  &  0.692604                          \\
NSVS 4161544  &  SPb  &  0.296  &  65  &  0.098316                     \\
OU Ser  &  SPb  &  0.173  &  50.47  &  0.667473                        \\
PY Vir  &  SPb  &  0.773  &  69.53  &  0.003462                        \\
RR Cen  &  DP  &  0.21  &  81  &  0.434061                             \\
RW Com  &  SPb  &  0.471  &  72.43  &  0.176115                        \\
RW Dor  &  SPb  &  0.63  &  77.2  &  0.026321                          \\
RZ Com  &  DP  &  0.46  &  86.8  &  0.171163                           \\
RZ Tau  &  DP  &  0.376  &  83.57  &  0.56057                          \\
SS Ari  &  SPp  &  0.308  &  77.31  &  0.093791                        \\
SW Lac  &  SPf  &  0.787  &  79.8  &  0.354653                         \\
SZ Hor  &  DP  &  0.47  &  80.36  &  0.140656                          \\
TT Cet  &  DP  &  0.39  &  82.25  &  0.08                              \\
TV Mus  &  DP  &  0.166  &  77.15  &  0.7421                           \\
TW Cet  &  SPf  &  0.75  &  81.18  &  0.06                             \\
TX Cnc  &  SPb  &  0.455  &  62.19  &  0.170047                        \\
TY UMa  &  DP  &  0.396  &  84.9  &  0.132607                          \\
TZ Boo  &  DP  &  0.207  &  85.45  &  0.531299                         \\
UV Lyn  &  SPb  &  0.372  &  67.6  &  0.188714                         \\
UX Eri  &  SPb  &  0.373  &  75.7  &  0.204404                         \\
V1073 Cyg  &  SPp  &  0.303  &  68.4  &  0.121013                      \\
V1123 Tau  &  SPp  &  0.279  &  74.01  &  0.295456                     \\
V1128 Tau  &  DP  &  0.534  &  86  &  0.133974                         \\
V1191 Cyg  &  DP  &  0.107  &  80.76  &  0.577083                      \\
V1918 Cyg  &  DP  &  0.278  &  80.54  &  0.305128                      \\
V345 Gem  &  DP  &  0.142  &  73.3  &  0.999713                        \\
V351 Peg  &  SPb  &  0.36  &  63  &  0.206535                          \\
V402 Aur  &  SPb  &  0.2  &  52.65  &  0.030596                        \\
V410 Aur  &  DP  &  0.144  &  81.9  &  0.319156                        \\
V524 Mon  &  SPb  &  0.476  &  80.47  &  0.080554                      \\
V535 Ara  &  DP  &  0.302  &  79.88  &  0.223899                       \\
V546 And  &  DP  &  0.254  &  86.07  &  0.300058                       \\
V592 Per  &  DP  &  0.389  &  89.6  &  0.544801                        \\
V728 Her  &  SPb  &  0.178  &  69.3  &  0.81                           \\
V757 Cen  &  SPb  &  0.671  &  69.31  &  0.144196                      \\
V776 Cas  &  SPb  &  0.13  &  55.4  &  0.645645                        \\
V829 Her  &  SPb  &  0.435  &  57.2  &  0.199746                       \\
V842 Her  &  DP  &  0.259  &  78.3  &  0.254273                        \\
V868 Mon  &  DP  &  0.373  &  83.67  &  0.459621                       \\
V870 Ara  &  DP  &  0.082  &  70  &  0.984479                          \\
V899 Her  &  SPb  &  0.566  &  68.72  &  0.235797                      \\
V921 Her  &  SPp  &  0.226  &  75.4  &  0.001433                       \\
VW LMi  &  SPb  &  0.423  &  78.1  &  0.504979                         \\
VY Sex  &  SPb  &  0.313  &  65.53  &  0.246474                        \\
W UMa  &  DP  &  0.49  &  86  &  0.099593                              \\
XX Sex  &  DP  &  0.1  &  74.88  &  0.476294                           \\
Y Sex  &  DP  &  0.195  &  79.12  &  0.596136                          \\
YY CrB  &  DP  &  0.232  &  79.5  &  0.228296                          \\
YY Eri  &  DP  &  0.454  &  82.12  &  0.050994                         \\
\end{longtable}

\begin{figure*}
\begin{center}
  \includegraphics[width=0.7\textwidth]{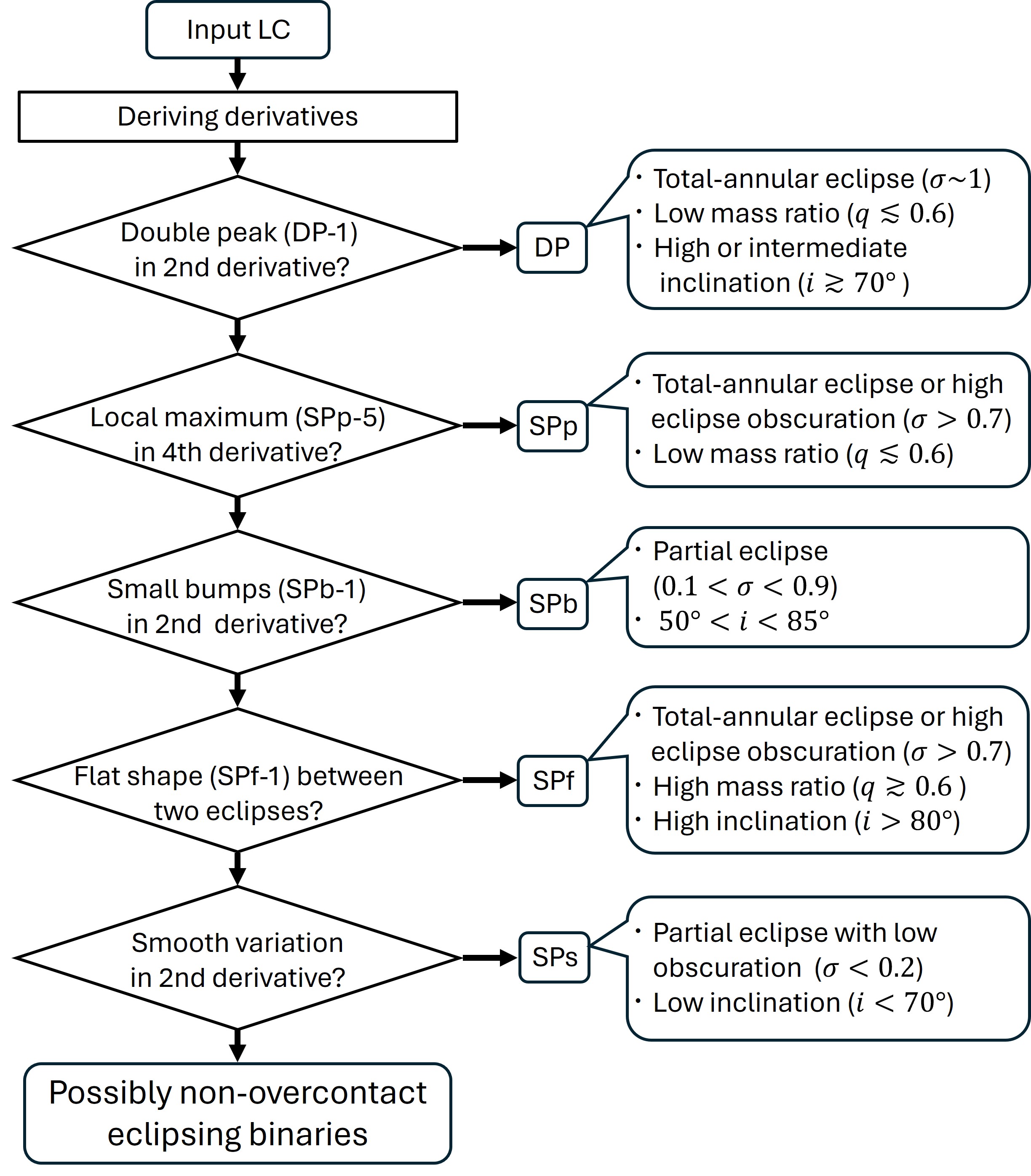}
 \end{center}
 \caption{Flowchart of classifying the derivatives of a LC. 
		  Diamonds denote decision points to classify according to the proposed five types. 
		  Dialogue balloons summarize the general properties of the five types. 
		  {Alt text: Flowchart describes criteria for classifying the LCs of overcontact eclipsing binaries. }
			\label{fig_flow}}
\end{figure*}
\section{Summary}\label{Sec_Summary}
We have proposed a new classification method for the LCs of overcontact eclipsing binaries by using their first to fourth derivatives. 
On the basis of morphological characteristics of their derivatives, we divided our synthesized sample LCs into five categories: DP, SPp, SPb, SPf, and SPs types. 
For each type, we also examined the statistical distributions of four key parameters (the mass ratio, orbital inclination, fill-out factor, and eclipse obscuration). 
Consequently, we found that each type has distinguishing physical properties. 
Figure \ref{fig_flow} presents a flowchart of the proposed classification method and also summarizes the general properties of the proposed types. 
When the shape of the derivatives makes it difficult to determine which of the two types is more appropriate, the system often exhibits intermediate properties of both types. 

In this classification method, overcontact eclipsing binaries with a low mass ratio and high orbital inclination, sufficient to cause total-annular eclipses, would be classified as DP type. 
Binaries with a low mass ratio but a relatively low inclination, resulting in total-annular eclipses or eclipses with a high obscuration, should be classified as SPp type. 
When binaries have a high mass ratio and high inclination, they should have large eclipse obscurations and be classified as SPf type. 
If binaries have a low inclination, they should show grazing eclipses and be classified as SPs type. 
Other binaries exhibiting partial eclipses are expected to be classified as SPb type.

The proposed classification method allows us to know general information about some physical parameters of overcontact eclipsing binaries. 
Such information should provide a basis for further detailed analyses of these systems. 
For example, our method can be helpful for setting initial parameters when modeling a LC. 
Moreover, recent surveys have discovered numerous eclipsing binaries. 
Currently, it is challenging to simultaneously obtain the properties of those binaries through precise analysis. 
However, applying the proposed classification also enables us to efficiently obtain general trends and physical insights about a large number of overcontact binary systems. 

\begin{ack}
This paper includes data collected by the TESS mission. Funding for the TESS mission is provided by the NASA’s Science Mission Directorate. 
This paper includes data collected by the Kepler mission and obtained from the MAST data archive at the Space Telescope Science Institute (STScI). 
Funding for the Kepler mission is provided by the NASA Science Mission Directorate. 
STScI is operated by the Association of Universities for Research in Astronomy, Inc., under NASA contract NAS 5–26555. 
\end{ack}


\begin{thebibliography}{16}
\expandafter\ifx\csname natexlab\endcsname\relax\def\natexlab#1{#1}\fi

\bibitem[{{Binnendijk}(1970)}]{Binnendijk1970-VA}
{Binnendijk}, L. 1970, Vistas in Astronomy, 12, 217

\bibitem[{{Conroy} {et~al.}(2020){Conroy}, {Kochoska}, {Hey}, {Pablo},
  {Hambleton}, {Jones}, {Giammarco}, {Abdul-Masih}, \&
  {Pr{\v{s}}a}}]{Conroy2020-ApJS}
{Conroy}, K.~E., {Kochoska}, A., {Hey}, D., {et~al.} 2020, \apjs, 250, 34

\bibitem[{{Gazeas} \& {Niarchos}(2006)}]{Gazeas2006-MNRAS}
{Gazeas}, K.~D. \& {Niarchos}, P.~G. 2006, \mnras, 370, L29

\bibitem[{{IJspeert} {et~al.}(2021){IJspeert}, {Tkachenko}, {Johnston},
  {Garcia}, {De Ridder}, {Van Reeth}, \& {Aerts}}]{IJspeert2021-AA}
{IJspeert}, L.~W., {Tkachenko}, A., {Johnston}, C., {et~al.} 2021, \aap, 652,
  A120

\bibitem[{{IJspeert} {et~al.}(2024){IJspeert}, {Tkachenko}, {Johnston},
  {Pr{\v{s}}a}, {Wells}, \& {Aerts}}]{IJspeert2024-AA}
{IJspeert}, L.~W., {Tkachenko}, A., {Johnston}, C., {et~al.} 2024, \aap, 685,
  A62

\bibitem[{{Kallrath} \& {Milone}(2009)}]{Kallrath2009-book}
{Kallrath}, J. \& {Milone}, E.~F. 2009, {Eclipsing Binary Stars: Modeling and
  Analysis}

\bibitem[{{Kopal}(1955)}]{Kopal1955-AnAp}
{Kopal}, Z. 1955, Annales d'Astrophysique, 18, 379

\bibitem[{{Kouzuma}(2023)}]{Kouzuma2023-ApJ}
{Kouzuma}, S. 2023, \apj, 958, 84

\bibitem[{{Latkovi{\'c}} {et~al.}(2021){Latkovi{\'c}}, {{\v{C}}eki}, \&
  {Lazarevi{\'c}}}]{Latkovic2021-ApJS}
{Latkovi{\'c}}, O., {{\v{C}}eki}, A., \& {Lazarevi{\'c}}, S. 2021, \apjs, 254,
  10

\bibitem[{{Lightkurve Collaboration} {et~al.}(2018){Lightkurve Collaboration},
  {Cardoso}, {Hedges}, {Gully-Santiago}, {Saunders}, {Cody}, {Barclay}, {Hall},
  {Sagear}, {Turtelboom}, {Zhang}, {Tzanidakis}, {Mighell}, {Coughlin}, {Bell},
  {Berta-Thompson}, {Williams}, {Dotson}, \& {Barentsen}}]{Lightkurve2018-code}
{Lightkurve Collaboration}, {Cardoso}, J.~V.~d.~M., {Hedges}, C., {et~al.}
  2018, {Lightkurve: Kepler and TESS time series analysis in Python},
  Astrophysics Source Code Library

\bibitem[{{Pr{\v{s}}a} {et~al.}(2016){Pr{\v{s}}a}, {Conroy}, {Horvat}, {Pablo},
  {Kochoska}, {Bloemen}, {Giammarco}, {Hambleton}, \&
  {Degroote}}]{Prsa2016-ApJS}
{Pr{\v{s}}a}, A., {Conroy}, K.~E., {Horvat}, M., {et~al.} 2016, \apjs, 227, 29

\bibitem[{{Rucinski}(1974)}]{Rucinski1974-AcA}
{Rucinski}, S.~M. 1974, \actaa, 24, 119

\bibitem[{{Samus} {et~al.}(2004){Samus}, {Durlevich}, \& {et
  al.}}]{Samus2004-catalogue}
{Samus}, N.~N., {Durlevich}, O.~V., \& {et al.} 2004, VizieR Online Data
  Catalog, 2250, 0

\bibitem[{{Smith}(1984)}]{Smith1984-QJRAS}
{Smith}, R.~C. 1984, \qjras, 25, 405

\bibitem[{{Webbink}(2003)}]{Webbink2003-ASPC}
{Webbink}, R.~F. 2003, in Astronomical Society of the Pacific Conference
  Series, Vol. 293, 3D Stellar Evolution, ed. S.~{Turcotte}, S.~C. {Keller}, \&
  R.~M. {Cavallo}, 76

\bibitem[{{Yamasaki}(1975)}]{Yamasaki1975-ApSS}
{Yamasaki}, A. 1975, \apss, 34, 413

\end{thebibliography}
\end{document}